\documentclass[prl,showpacs,amssymb,floatfix,twocolumn]{revtex4}
\usepackage{graphicx}
\usepackage{epsfig}
\usepackage{dcolumn}
\usepackage{bm}
\usepackage{times}
\usepackage{amsmath}
\usepackage{amssymb}
\usepackage{mathrsfs}

\def\(({\left(}
\def\)){\right)}
\def\[[{\left[}
\def\]]{\right]}

\newcommand{\be}{\begin{equation}}
\newcommand{\ee}{\end{equation}}
\newcommand{\bea}{\begin{eqnarray}}
\newcommand{\eea}{\end{eqnarray}}

\begin{document}

\title{Jamming versus Glass Transitions}

\author {Romain Mari$^1$, Florent Krzakala$^{2}$, Jorge Kurchan$^1$ }
\affiliation{ $^1$ CNRS; ESPCI, 10 rue Vauquelin, UMR 7636 PMMH,
  Paris,
  France 75005,\\
  $^2$ CNRS; ESPCI, 10 rue Vauquelin, UMR 7083 Gulliver, Paris, France
  75005
}

\begin{abstract}
  Recent ideas based on the properties of assemblies of frictionless
  particles in mechanical equilibrium provide a perspective of
  amorphous systems different from that offered by the traditional
  approach originating in liquid theory.  The relation, if any,
  between these two points of view, and the relevance of the former to
  the glass phase, has been difficult to ascertain.  In this paper we
  introduce a model for which {\em both} theories apply strictly: it
  exhibits on the one hand an ideal glass transition and on the other
  `jamming' features (fragility, soft modes) virtually identical to
  that of real systems.  This allows us to disentangle the two
  physical phenomena.
\end{abstract}
  

\maketitle

The traditional way to introduce the glass transition is to start with
a liquid of, say, hard particles at low pressures \cite{foot2}, and to
consider a slow compression.  At a given point, the viscosity
increases dramatically, the dynamics becomes sluggish and the system
falls out of equilibrium.  Slower compression protocols push the
equilibrium regime further and make the transition sharper, and one
conceives of a limit of infinitely slow annealing in which one
recognizes (perhaps) a true thermodynamic change of state.  An
argument in favor of a thermodynamic transition was given years ago by
Kauzmann, who interpreted it as a consequence of the liquid running
out of configurational entropy.  Although the possibility of proving
the occurrence of such an ideal glass transition for a real system
seems remote, there is a family of models (or approximations), for
which this picture holds strictly, within the so-called `Random First
Order' (RFO) scenario \cite{Wolynes_RFO,Review}, a Mean-Field theory allowing
for a complete analytic analysis.

An apparently unrelated set of ideas comes from considering amorphous
assemblies of hard, frictionless particles in mechanical equilibrium.
In general, such systems can be hypostatic, hyperstatic or isostatic,
depending on whether the set of contacts yields a number of conditions
smaller, larger, or precisely equal to the number of degrees of
freedom -- just like a table with two, four or three legs touching a
floor.  A polydisperse system of spheres will with probability one be
isostatic~\cite{Moukarzel,Roux_isostatic,Witten_isostatic}, much in
the same way that a table with many legs will only have three touching
on a rough floor.  Because breaking one single contact already
destabilizes an isostatic system, one can then argue that such systems
are {\em marginal}~\cite{Ohern,Witten_isostatic,BritoWyart}: they have
large responses and a spectrum of vibrations with finite density of
very low frequency `soft' modes. One is then in the presence of a
``Jammed'' configuration that is {\em critical}, with diverging
lengths, nontrivial exponents, etc~\cite{Ohern}.

A natural question is whether this critical ``Jamming'' phenomenon is
in some way a finite-dimensional (beyond mean-field) manifestation of
the glass transition~\cite{Ohern} or in other words, whether the
diverging length associated with the soft modes of an amorphous
packing is a manifestation of the order underlying the glass
transition in finite dimensions. In this paper we propose to answer
such questions by constructing a family of models having both theories
written side-by-side with a `J-point' isostatic equilibrium with a
spectrum of soft modes virtually identical to that of systems of hard
spheres and that are by construction mean-field models with a `Random
First Order' behavior~\cite{Wolynes_RFO,Review}.
\begin{figure}
  \centering
  \includegraphics[width=0.8\columnwidth]{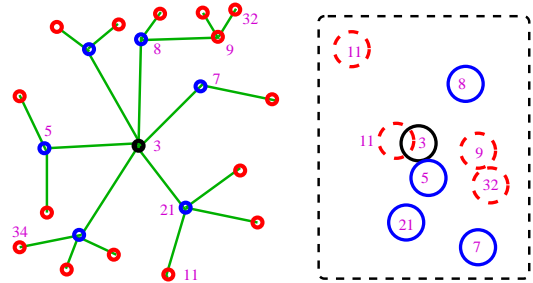}
  \caption{The model: Particles evolve in a $d$-dimensional space
    (right), but are only able to `see' a pre-established subset of the
    rest. Which particle interacts with which is encoded in the
    quenched graph (left). 
}
  \label{graph}
\vspace{-0.4cm}
\end{figure}

\paragraph*{The model ---}
Consider the usual hard-sphere model where $N$ non overlapping spheres
live in a cube of dimension $d$, with periodic boundary
conditions. Following an approach introduced in \cite{KK}, we now
obtain a mean field version of this model: the idea is to let each
sphere interact {\it only} with a set of $z$ other ones, all the rest
being {\it transparent} to it. The set of particles that interact with
a given sphere is chosen once and for all.  If $z=N-1$ for each
particles, we recover the original problem, but we shall instead work
with $z$ finite. The model is completely defined by a quenched regular
random graph (see FIG.\ref{graph}) with nodes labeled by the particle
number and bonds that denote interaction.

This model presents a number of advantages: first, since interactions
are set through a random tree-like graph ---or Bethe-Lattice--- it is
by construction a mean-field glass model that can be studied exactly
with the cavity method~\cite{Cavity} which is an extension of the
well-known work of Bethe and Peierls on trees.  Secondly, our model
can be seen as a constraint satisfaction problem (CSP) defined on a
random graph, where the constraint on each node is that the position
within the cube of the sphere on that node be such that it does not
overlap with any of the other $z$ interacting spheres (that are linked
on the graph). This model bridges the gap between the field of random
discrete CSP such as the coloring problem ~\cite{coloring1,coloring2}
and disordered hard-sphere packings~\cite{KK}.  Our approach thus
provides an original way to attack packing problems within mean field
theory. In this paper, we take the first steps in this direction by
studying numerically the model in $d=2$ and restrict ourselves to the
question of determining the differences between the Jamming and the
glass transition.

\begin{figure}
  \centering \includegraphics[width=0.9\columnwidth]{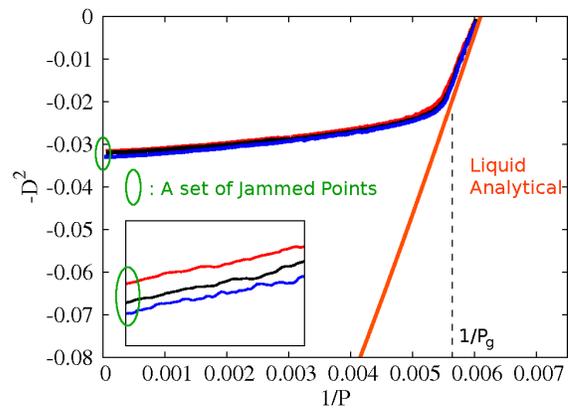}
  \caption{Volume  versus inverse pressure for a system of $N=4000$
    particles with $z=100$ and 9\% polydispersity. The choice of variables for the  axes is to stress the analogy with the more usual energy-temperature annealing plots. The annealing curves follow the liquid 
    analytical solution at low pressures,  but break away at the glass transition pressure $P_g$, which is clearly visible.
    }
  \label{cooling}
\end{figure}

\paragraph*{Glassy behavior ---} The low pressure/low density
``liquid'' state can be discussed easily within the cavity method. It
just corresponds to considering the model on a tree in the usual
Bethe-Peierls way while ignoring long range correlations. A
straightforward computation shows that the entropy of the liquid state
on a random graph of connectivity $z$ is $S_{liquid}=z \log{(1-\pi
  D^2)}/2$, where $D$ is the diameter of the
spheres~\cite{MPTZ}. Using a pressure $-P$ conjugated to the sphere
volume~\cite{KK} $V=D^2$ this yields the equation of state in the
liquid phase $P=(z\pi/2)(1-\pi D^2)^{-1}$. However, this equation is
inconsistent at large pressures: indeed for $P=\infty$ it yields a
limiting value independent of the connectivity $z$, which is
unphysical, and where each sphere is in contact with all its
neighbors; this requires not only the number of contacts to be much
larger than the isostatic value, but also the presence of a
$2-$coloring of the graph \cite{KK} and thus violates rigorous results
in graph theory: this proves that a phase transition {\it must}
intervene at a finite pressure value. It is indeed possible to prove
within the cavity method that the liquid phase is unstable towards a
glass phase at some finite pressure and ongoing analytic work in this
direction shows that the system has a glass phase~\cite{MPTZ} of the
Random First Order kind for $z$ sufficiently large, as usual with
frustrated models on such structures
\cite{Cavity,coloring1,coloring2,MPTZ,biroli,flo,Semerjian}. This is
illustrated on Fig.\ref{cooling} in a system with a slight
polydispersity \cite{poly} (in which case one should compute the averaged
$S_{liquid}=z \overline{\log{(1-\pi (R_1+R_2)^2)}/2}$ over the
distribution of radii $P(R)$): a clear signal of glass transition with
its characteristic compression rate dependence is observed.

\begin{figure*}[t]
\begin{center}
\hspace{-1.7cm}
\includegraphics[width=6.4cm]{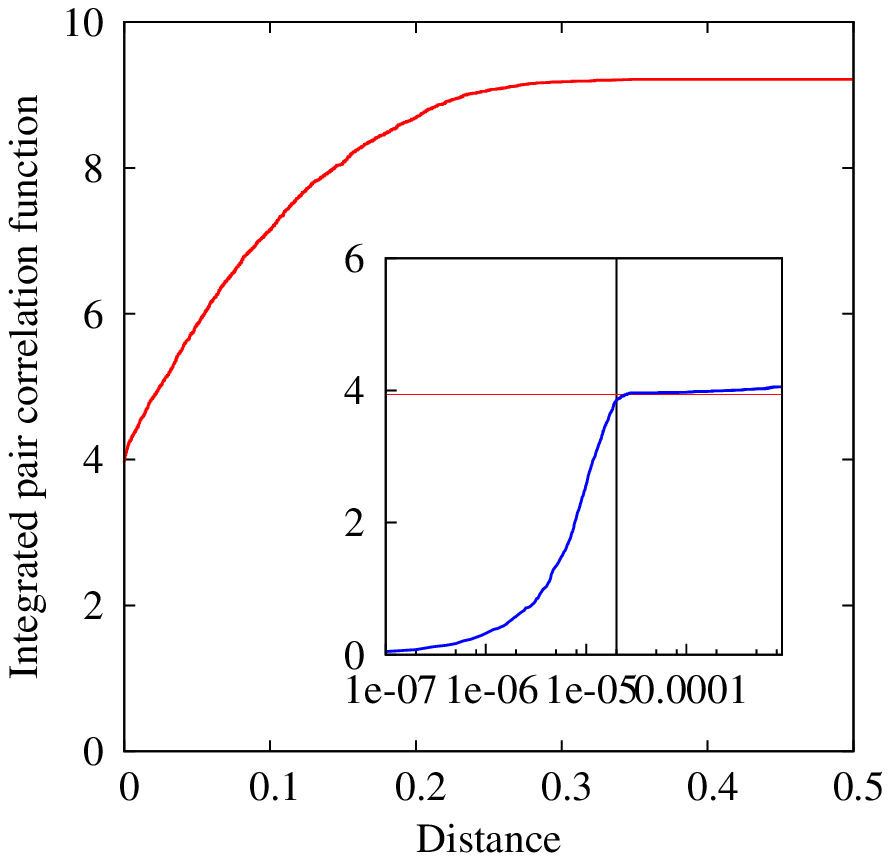}
\hspace{0.15cm}
\includegraphics[width=6.4cm]{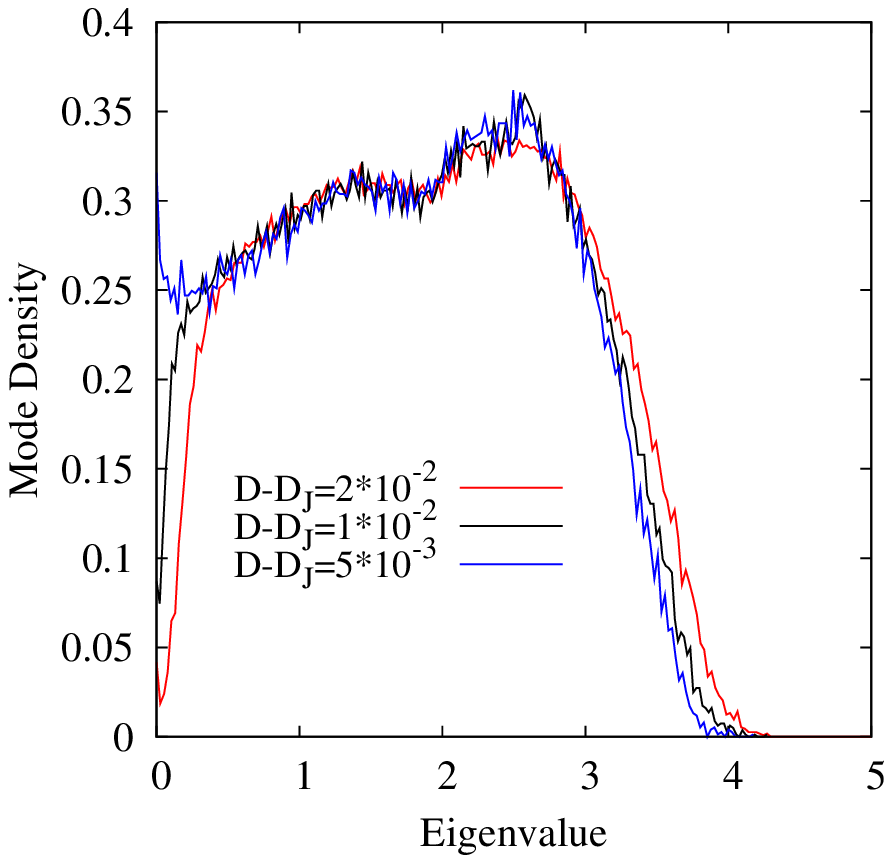}
\hspace{-1.5cm}
\includegraphics[width=6.4cm]{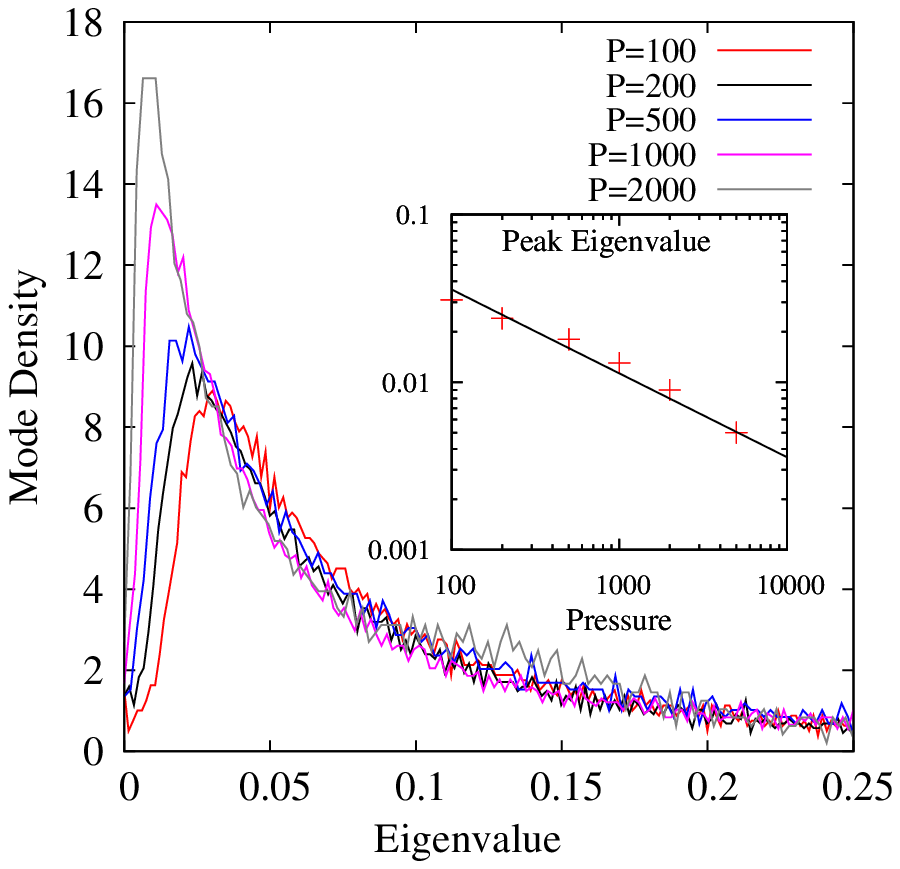}
\caption{{\bf Left:} Integrated pair correlation function restricted
  to the collectively jammed core. The average number of neighbors is
  four, as one expects from an isostatic system in two
  dimensions. The inset shows a zoom in the small distance region, 
  the horizontal line is the isostaticity condition, while the vertical 
  line is the gradient descent step used to jam the system. 
  {\bf Center:} Vibration modes with small overlaps (here
  noted $D-D_J$, difference between actual diameter of the particles 
  and diameter particles should have to be at jamming).  The resulting 
  spectrum is almost identical to that
  of a particle system (cf FIG. 1 of ref. \cite{Silbert}). {\bf
  Right:} Hard particles: vibration modes of the isostatic core,
  rescaled with the pressure. In the inset, gap in the spectrum
  vs. pressure scales as $P^{-1/2}$.  The curves are virtually
  identical to those of a finite dimensional system, as obtained by
  Brito and Wyart \cite{BritoWyart}. These curves correspond to
  a configuration with N=1000 and z=13.
  \label{modes}}
\end{center}
\end{figure*}

\paragraph*{Jamming ---} We now turn to jammed configurations obtained
following the procedure of \cite{Ohern}: starting from small particles
we `inflate' them infinitesimally, and adjust their positions to
eliminate any (infinitesimal) overlap, until a jammed state is
obtained. At very large pressure, particles are found to be either locally 
blocked by their neighbors (in which case they have at least three contacts), 
or are `rattlers' that can be displaced without moving the 
others ~\cite{Chayes}. The locally blocked particles have a subset 
that cannot be displaced even with collective rearrangements 
-- except global translations. For our polydisperse system, this 
subset turns out not to have redundant contacts, and hence constitutes 
an {\em isostatic core} (see left panel of FIG. \ref{modes}).  We 
find many states with slightly different densities, all of which 
are isostatic and have indiscernible properties.

The jammed configurations have a spectrum of modes that is strikingly
similar to those found in the ordinary finite dimensional
system~\cite{Silbert,BritoWyart,Saarloos}.  In FIG. \ref{modes} we
show the vibration modes obtained by considering a soft potential and
a small superposition of particles that follows from inflating the
particles slightly beyond the jammed configuration, as in
Ref. \cite{Silbert}.  A second, perhaps more satisfactory way to study
the vibrations of a hard sphere ensemble is to consider the very high
pressure dynamics of hard particles around a jammed state, and
computing the displacement correlations $\langle A_{ij} \rangle$ with
$A_{ij}= x_i(t) x_j(t) - \langle x_i \rangle \langle x_j \rangle $
where $\langle \bullet \rangle$ denotes average over a time that is
long yet insufficient for escaping the vicinity of the jammed
configuration. The displacements fall into two classes: the
many-particle vibrations that scale as $P^{-1}$, and the rattlers that
do not, because the cages are roughly independent of the pressure.
The role of the frequencies in a potential is played by the
eigenvalues of $P^{-1} A^{-1/2}$. The spectrum of these eigenvalues is
shown in FIG. \ref{modes}: at infinite pressure there is a
proliferation of soft modes, but as the volume fraction $\varphi$ is
lowered, and the pressure becomes finite, the system interacts, due to
collisions and rattling, with more particles than what is imposed by
the isostaticity condition, and the isostaticity-related soft modes
start disappearing. Again, this is in perfect agreement with the usual
particle systems \cite{BritoWyart}.

\begin{figure}
  \centering
  \includegraphics[width=0.75\columnwidth]{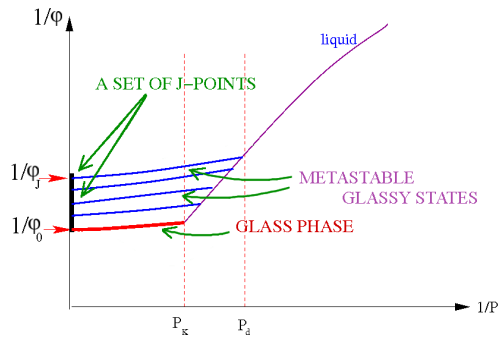}
  \caption{A sketch of the density($\varphi$)-pressure($P$) plane in RFO
    models (see e.g. \cite{Review,flo,KK}). Dynamic and static
    transition pressures are $P_d$ and $P_K$.  The glass phase and the
    `jamming line' $(P=\infty, \varphi_o<\varphi<\varphi_J)$ are clearly
    distinct.  The isostaticity-related quantity $\chi_{EA}$ is finite
    within a state if $P$ is finite (except perhaps for a small effect
    of acoustic modes in $d=2$).  In particular it is finite within
    the ideal glass state, while it is infinite on the jamming line.
    Everywhere in the trapezoidal region delimited by the equilibrium
    line $0<P<P_d$, the threshold level, and the jamming line the
    activation time diverges (at times exponential in $N$), and with
    it the four-point function $\chi_4$. In finite dimensions $\chi_4$
    would only diverge strictly at the jamming line (at finite times)
    and close to the glass line (at times comparable to the activation
    time). }
  \label{phase}
\end{figure}

\paragraph*{Jamming versus Glass ---} 
Having discussed the isostatic packings with the standard $J$-point
phenomenology \cite{Ohern,Silbert,Barrat,BritoWyart},  as well as the
mean field glassy nature of our model, we can now distinguish the
`Jammed' and the `Glass' features as follows (FIG. \ref{phase}). There
is a liquid equilibrium line, which terminates at the equilibrium
glass transition pressure $P_K$.  Below $P_K$ the Gibbs measure is
dominated by a few of the deepest glassy states \cite{Zampouno}. There
are also metastable ones (indicated by the horizontal lines), which
are stable (only) within the mean-field picture. As in any system with
a dynamic transition, any compression process ending at infinite
pressure leads the system into a metastable state. The `jamming line'
$\{P=\infty \; , \; \varphi_o<\varphi<\varphi_J\}$ is the set of such (out of
equilibrium) blocked configurations. The slower the compression, the
denser the target state is, and one needs an infinitely slow process
to reach $\varphi_0$.  In particular, we denote the $J$ point $\varphi_J$ as
the result of the fastest compression starting from a random
configuration~\cite{Ohern}, which we expect to be different from the
deepest level $\varphi_o$ -- just as a quench to zero temperature does
not terminate in the ground state of any complex system. In a finite
dimensional system, only two lines in the phase diagram are stable,
apart from the liquid state: the jamming line (at infinite pressure)
and the equilibrium glass phase -- if it exists at all.  Their only
common point is $(\varphi_o,P=\infty)$.  

A measure of the criticality of a state is the staggered displacement
$\langle \sum_i \vec \xi_i \vec x_i \rangle/h$ produced by random
forces $\vec f_i=h \vec \xi_i$, with $\vec \xi_i$ random unit
vectors.  This corresponds, in the spin-glass literature, to the
quantity $(1-q_{EA})/T$. It is proportional to $\chi_{EA}= P^2
\sum_i (\langle x^2_i(t)\rangle -\langle x_i\rangle^2)=P^2\text{Tr}(\langle A \rangle )$. If $\omega$ are the eigenvalues of $P^{-1}A^{-1/2}$ and $Q(\omega)$ 
the density of modes, then 
$\chi_{EA}\sim \int \text{d}\omega Q(\omega)/ \omega^2$. Due to the
soft modes~\cite{BritoWyart}, this quantity diverges on approaching
the jamming line when normalised this way, see inset of FIG.
\ref{modes}. A perhaps more standard measure is the function
$\chi_4(t,t')=\sum_{ij} A_{ij}(t)A_{ij}(t')$. It has a maximum which
diverges as the system approaches {\em either} the jamming points 
(because of the soft modes) or the equilibrium glass line, because of
the divergence of activation time (i.e. the time needed to overcome a 
free-energy barrier).

Thinking in terms of {\em landscape}~\cite{KK}, the picture that
emerges is one of a system with many glassy states, separated by high and  wide
free-energy barriers, as usual in RFO models. Fragility and soft modes
are then a property of the (infinite pressure) jammed configurations
{\it within} each of these states, but states are not marginal in the
coarse grained view obtained as soon as temperature is nonzero and the
pressure finite (at least in all but the more superficial `threshold'
levels).  This separation between large valleys with critical bottom
is consistent with the fact that assemblies of frictionless particles
at zero temperature are at the same time fragile (even  slight shear
stresses make them creep some amount) while still capable of resisting
a shear stress proportional to the pressure without flowing
continuously (they have {\em internal friction} \cite{friction}.)

\begin{figure}
\hspace{-1.0cm} \includegraphics[width=0.75\columnwidth]{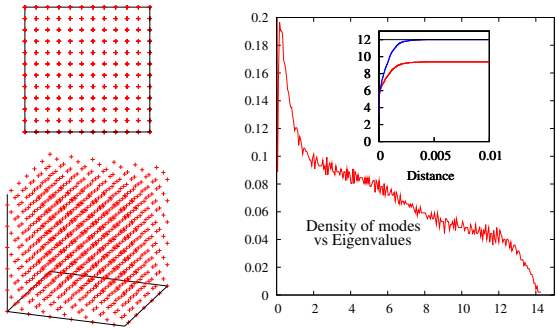}
  \caption{{\bf Left:} A jammed configuration in the polydisperse
    crystal (N=864).  The distortion of the crystalline order due to
    polydispersity is extremely weak. {\bf Right:} Spectrum of normal
    modes of the polydisperse crystal, obtained in the same way as FIG.3, 
    central panel.  There are many soft modes (To be compared with 
    FIG. \ref{modes}). In inset: integrated pair correlation function, 
    including and excluding rattlers: the number of contacts of non-rattlers 
    is $\approx 6$.
    \label{crystal}}
\end{figure}
\paragraph{Jamming a Crystal---} To unambiguously demonstrate that a
large stable state with an internal `fragile' structure is possible in
finite dimensions, and that the Jamming point is of different nature
than that of the glass transition, let us finally consider an ordinary
FCC crystal of spherical particles with very small (.003\%)
polydispersity. The crystalline order is hardly affected (see
FIG. \ref{crystal}), and polydispersity becomes irrelevant at finite
pressures $P$. However, at $P=\infty$ the system is isostatic, and,
not surprisingly, has a spectrum of soft modes even richer than that
of an amorphous packing.
In an amorphous solid, each one of the many equilibrium glass states
plays a similar role to the one of the crystal state above, their very
high pressure jamming properties being juxtaposed with the underlying
long-range glass order, already established at finite pressure.

\paragraph*{Conclusion ---} 
In this paper we confront two different visions of amorphous systems:
a glassy solid state with order characterized by a permanent,
amorphous modulation of density -- not unlike a crystal or a
quasi-crystal -- and a jamming situation brought about by chains of
force associated with actual contact between hard particles. These two
phenomena may coexist but are distinct.  The fact that a mean-field
model reproduces both the glass transition and J-point criticality in
a separate way suggests that we abandon the idea that the former is
some kind of finite-dimensional realization of the latter.  The
mean-field picture {\em just above the glass transition pressure} is
one of large basins separated by high barriers, without an excess of
truly zero frequency modes except perhaps at the least deep states.
The isostaticity-related marginality of the jammed configurations
appears only at very large pressures, deep within a basin. This local
criticality combined with absence of criticality `in the large' is
attested by the paradoxical fact that amorphous matter is fragile to
small stresses but may still sustain extensive stresses without
flowing, a fact that can be understood easily in the example of the
polydisperse crystal.  Perhaps the most evident manifestation of the
different nature of the jamming ($P=\infty$) and the glass lines is
the fact that, although lengths such as associated to four-point
function $\chi_4$ diverge in both, the growth is astronomically slower
approaching the glass line from the liquid phase, than approaching the
jamming line at $P=\infty$ \cite{olivier}. 

In conclusion, we have introduced a set of new models that can be
studied analytically and numerically and that provide non-trivial
connections between different fields. We hope this will be useful to
attack the glass and granular problems analytically in the future.

\acknowledgments
We wish to thank J. N. Roux for useful clarifications, and
M. M\'ezard, G. Parisi, M. Tarzia and F. Zamponi
for discussing with us their unpublished results \cite{MPTZ}.


\begin{thebibliography}{30}
\bibitem{foot2} Throughout this paper we consider that particles are
  thermally agitated, so that even when they are not permanently in
  contact they interact via collisions. Temperature thus plays a role.
\end{thebibliography}
\end{document}